\documentclass[11pt]{article}
\pdfoutput=1
\usepackage{graphicx,amssymb,amsmath,mathrsfs,array,multirow}  

\usepackage{caption,subcaption}
\usepackage[titletoc,toc,title]{appendix}
\usepackage{hyperref,color}
\definecolor{title}{RGB}{42,111,151}
\begin{document}  
 
\vskip 30pt

\begin{center}
{\Large \bf
Implications  of the CMS search for $W_R$ on Grand Unification
\\ }
\vskip 1cm
\renewcommand{\thefootnote}{\fnsymbol{footnote}}
Triparno Bandyopadhyay$^1$\footnote{email: gondogolegogol@gmail.com},
Biswajoy Brahmachari$^2$\footnote{email: biswa.brahmac@gmail.com},
and Amitava Raychaudhuri$^1$\footnote{email: palitprof@gmail.com}\\
\vskip 1cm
(1) Department of Physics, 
University of Calcutta,\\ 92 Acharya Prafulla Chandra  Road, 
Kolkata 700009, India
\vskip .5cm
(2) Department of Physics, 
Vidyasagar Evening College, \\
39 Sankar Ghosh Lane, Kolkata 700006, India\\
\end{center}
\vskip 2cm
\thispagestyle{empty}
\begin{center}
\underbar{\bf Abstract}
\end{center}

The CMS experiment at the Large Hadron Collider has reported a
2.8$\sigma$ excess in the $(2e)(2jets)$ channel around 2.1 TeV.
Interpretation of this data is reconsidered in terms of the
production of a right-handed weak gauge boson, $W_R$, of the
left-right symmetric model and in an $SO(10)$ grand unified
theory abiding by the Extended Survival Hypothesis.  The
left-right symmetric model can be consistent with this excess if
(a) the heavy right-handed neutrino has a mass near $W_R$, or (b)
if $g_L \neq g_R$, or (c) the right-handed CKM matrix is
nontrivial.  Combinations of the above possibilities are also
viable.  A $W_R$ with a mass in the TeV region if embedded in
$SO(10)$ is not compatible with $g_L = g_R$. Rather, it implies
$0.64 \leq g_R/g_L \leq 0.78$.  Further, a unique
symmetry-breaking route -- the order being left-right discrete
symmetry breaking first, followed by $SU(4)_C$ and finally
$SU(2)_R$ -- to the standard model is picked out. The $L
\leftrightarrow R$ discrete symmetry has to be broken at  around
$10^{16}$ GeV.  The grand unification scale is pushed to
$10^{18}$ GeV making the detection of proton decay in ongoing
searches rather unlikely.  The $SU(4)_C$ breaking scale can be at
its allowed lower limit of $10^6$ GeV so that  $n - \bar{n}$
oscillation or flavour changing processes such as $K_L
\rightarrow \mu e$  and $B_{d,s} \rightarrow \mu e$ may be
detectable.  The Higgs scalar multiplets  responsible for
$SO(10)$ symmetry breaking at various stages are uniquely
identified  so long as one adheres to a minimalist principle.  We
also remark, {\em en passant}, about a partially unified
Pati-Salam model.

\vskip 20pt

\noindent



\renewcommand{\thesection}{\Roman{section}} 
\setcounter{footnote}{0} 
\renewcommand{\thefootnote}{\arabic{footnote}} 
\noindent


\section{Introduction}

The discovery of the Higgs boson at the Large Hadron Collider (LHC)     
at CERN is a major milestone of the successes of the standard model     
(SM) of particle physics. Indeed, with all the quarks and leptons and   
force carriers of the SM now detected and the source of spontaneous     
symmetry breaking identified there is a well-deserved sense of          
satisfaction. Nonetheless, there is a widely shared expectation that    
there is new physics which may be around the corner and within  
striking range of the LHC. The shortcomings of the standard model       
are well-known. There is no candidate for dark matter in the SM. The    
neutrino is massless in the model but experiments indicate otherwise.   
At the same time the utter smallness of this mass is itself a mystery.  
Neither is there any explanation of the matter-antimatter asymmetry     
seen in the Universe. Besides, the lightness of the Higgs boson remains 
an enigma if there is no physics between the electroweak and Planck     
scales.

Of the several alternatives of beyond the standard model
extensions, the one on which we focus in this work is the
left-right symmetric (LRS) model \cite{LRS1, PS, LRS2, LRS3} and its
embedding within a grand unified theory (GUT). Here parity is a
symmetry of the theory which is spontaneously broken resulting in
the observed left-handed weak interactions. The left-right
symmetric model is based on the gauge group $SU(2)_L \times
SU(2)_R \times U(1)_{B-L}$ and has a natural embedding in the
$SU(4)_C \times SU(2)_L \times SU(2)_R$ Pati-Salam model
\cite{PS} which unifies quarks and leptons in an $SU(4)_{C}$
symmetry. The Pati-Salam symmetry is a subgroup of $SO(10)$
\cite{SO10a, SO10b}. These extensions of the standard model
provide avenues for the amelioration of several of its
shortcomings alluded to earlier.

The tell-tale signature of the LRS model would be  observation
of the $W_R$. At the LHC the CMS collaboration has searched for
the on-shell production of a right-handed charged gauge boson
\cite{cms} using the process\footnote{Earlier searches at the LHC
for the $W_R$ can be found in \cite {olderC, olderA}.}:
\begin{equation}
    pp\rightarrow W_R \rightarrow  2j+ll \;\;.
\label{KSmode}
\end{equation}
In the above  $l$ stands for a charged lepton, and $j$ represents
a hadronic jet. 

The CMS collaboration has examined the implication of its
findings in the context of a left-right symmetric model where the
left and right gauge couplings are equal ($g_L = g_R$) and also
the $W_R$ coupling to a charged lepton,  $l$, and its associated
right-handed neutrino, $N_l$, is diagonal with no leptonic
mixing,\footnote{The existence of three right-handed neutrinos --
$N_e, N_\mu$ and $N_\tau$ -- is acknowledged.} (i.e., $V_{N_ll} =
1$).  In the $l = e$
channel the data shows a 2.8$\sigma$ excess near 2.1 TeV. Also,
regions in the $M_{N_l} - M_{W_R}$ plane disfavoured by the data,
within an LRS theory with $g_L=g_R$, have been exhibited. After
production, the $W_R$ decays through $W_R \rightarrow l N_l$ in
the first stage. An associated signal of this process will be a
peak at $M_{N_l}$ in one of the $ljj$ invariant mass
combinations. CMS has not observed the latter. One
possibility may be that the produced $N_l$ has a substantial
coupling to the $\tau$-lepton \cite{other1,  dibosonTH2} -- $V_{N_l\tau}$ is
not small. Here we keep $M_{N_l}$ as a parameter of the model.

Within the LRS model there is room to admit the possibility       
of $g_L \neq g_R$. Interpretation of the CMS result in the presence     
of such a coupling asymmetry has also been taken up \cite{utpal,        
othergut} keeping $M_{N_l} = M_{W_R}/2$ and the implications for
grand unification and baryogenesis explored. In \cite{utpal} the
coupling parameter $V_{N_ll}$ is also allowed to deviate from unity.
Other interpretations of the excess have also appeared, for
example, in \cite{other1} - \cite{other4}.

In a left-right symmetric model emerging from a grand unified theory,   
such as $SO(10)$, one has a discrete symmetry $SU(2)_L \leftrightarrow  
SU(2)_R$ -- referred to as D-parity \cite{Dparity} -- which sets $g_L   
= g_R$. Both D-parity and $SU(2)_R$ are broken during the descent of    
the GUT to the standard model, the first making the coupling constants  
unequal and the second resulting in a massive $W_R$. The possibility    
that the energy scale of breaking of D-parity is different from
that of $SU(2)_R$ breaking is admissible  and well-examined
\cite{so10-many}. The difference between these scales and the
particle content of the theory controls the extent to which $g_L
\neq g_R$.

In this work we consider the different options of $SO(10)$
symmetry breaking. It is shown that a light $W_R$ goes
hand-in-hand with the breaking of D-parity at a scale around
$10^{16}$ GeV, immediately excluding the possibility of $g_L =
g_R$.  D-parity breaking at such an energy is usually considered
a desirable feature for getting rid of unwanted  topological
defects such as domain walls \cite{domain1} and  accounting for
the baryon asymmetry of the Universe \cite{kuzmin}.  The
other symmetries that are broken in the passage to the standard
model are the $SU(4)_C$ and $SU(2)_R$ of the Pati-Salam (PS)
model. The stepwise breaking of these symmetries and the order of
their energy scales have many variants. There are also a variety
of options for the scalar multiplets which are used to trigger
the spontaneous symmetry breaking at the different stages. We
take a minimalist position of (a) not including any scalar fields
beyond the ones that are essential for symmetry breaking, and
also (b) impose the Extended Survival Hypothesis (ESH)
corresponding to minimal fine-tuning to keep no light extra
scalars. With these twin requirements we  observe that only a single
symmetry-breaking route -- the one in which the order of symmetry
breaking is first D-parity, then $SU(4)_C$, and finally $SU(2)_R$
-- can accommodate a light $M_{W_R}$. We find that one must have
$0.64 \leq g_R/g_L \leq 0.78$.

The paper is divided as follows. In the following section we give
details of the CMS result \cite{cms} which are relevant for our
discussion within the context of the left-right symmetric model.
In the next section we elaborate on the GUT symmetry-breaking
chains, the extended survival hypothesis for light scalars, and
coupling constant evolution relations.  Next we briefly note the
implications of coupling constant unification within the
Pati-Salam and $SO(10)$ models.  The results which emerge for the
different routes of descent of $SO(10)$ to the SM are presented
in the next two sections. We end with our conclusions.


\section{CMS $W_R$ search  result and the Left-Right Symmetric model}

The results of the CMS collaboration for the search for a
$W_R$-boson that we use \cite{cms} are based on the LHC run at
$\sqrt{s}$ = 8 TeV with an integrated luminosity of $19.7\text{
fb}^{-1}$. The focus is on the production of a $W_R$ which then
decays to a charged lepton ($l$) and a right-handed heavy
neutrino ($N_l$), both of which are on-shell.  The $N_l$
undergoes a three-body decay to a charged lepton ($l$) and a pair
of  quarks which manifest as hadronic jets ($2j$), the process
being mediated by a $W_R$. CMS examines the $(2l)(2j)$ data
within the framework of an LRS model with $g_L = g_R$ and
presents exclusion regions in the $M_{W_R}-M_{N_l}$
plane\footnote{An alternate explanation of the excess in the data
could be in terms of a charged Higgs boson of the LRS model.}.
Interpreting the four-object final state mass as that of a $W_R$
CMS presents, in the supplementary material of \cite{cms}, the
95\% CL exclusion limits for the observed and expected
$\sigma(pp\rightarrow W_R)\times BR(W_R\rightarrow lljj) \equiv
\sigma BR$ as functions of $M_{W_R}$ for several $M_{N_l}$. From
the data \cite{cms} one finds that in the electron channel,
irrespective of the value of $r=\frac{M_{N_e}}{M_{W_R}}$, $\sigma
BR_O$ (observed) exceeds twice the expected exclusion limit 
($\sigma BR_E$) for $1.8\lesssim M_{W_R}
\lesssim 2.4$ TeV.  This excess is about $\sim 2.8\sigma$ around
2.1 TeV.  Though not large enough for a firm conclusion, this can
be taken as a tentative hint for a $W_R$, and if this is correct,
one can expect confirmation in the new run of the LHC at
$\sqrt{s} = 13$ TeV. The CMS collaboration notes that this excess
is not consistent with the LRS model with $g_L = g_R, ~r = 0.5$
and no leptonic mixing.  As we stress later, relaxing these conditions --
e.g., $r =$ 0.5 -- can make the results agree with the left-right
symmetric model. No such excess is seen in the $(2\mu)(2j)$ mode.

\subsection{A $W_R$ signal?}
\label{sec:WR}

\begin{figure}[h!]
    \centering
    \includegraphics[width=.4\textwidth]{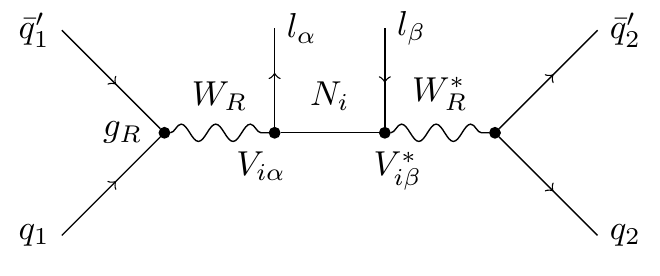}
    \caption{\em Feynman diagram of the process under discussion
with the right-handed CKM-like mixing matrix taken in a general,
non-diagonal, form.  The CMS excess corresponds to $l_\alpha =
l_\beta = e$.}
    \label{process}
\end{figure}

Fig. \ref{process} shows the Feynman diagram for $W_R$ production       
and its decay in the channels under consideration \cite{keung}. Note that the        
production of the $W_R$ will be suppressed compared to that of a        
left-handed $W$ boson of the same mass by a factor $\eta^2$, where      
$\eta = (g_R/g_L)$. The contribution from this diagram is determined    
by $S^2$ where $S \equiv \eta |V_{N_ee}|^2$. Neglecting the       
masses of the final state quarks and the charged lepton, the branching  
ratio of the three-body decay of $N_e$, which we have calculated, is    
proportional to $(1 - r^2)^2 (2 + r^2)$, where, as noted earlier, $r    
= M_{N_e}/M_{W_R}$. A clinching evidence of this process would then     
be a peak in the $(2e)(2j)$ invariant mass at $M_{W_R}$ -- for which    
there is already a hint -- along with another around $M_{N_e}$ in the   
invariant mass of one of the two $e (2j)$ combinations in every event.  
The absence of the latter in the data could be indicative of
more than one $N_R$ state being involved and further the coupling
of these neutrinos to a $\tau$-lepton with non-negligible strength
\cite{other1, dibosonTH2, dibosonTH1}. Subsequent leptonic decays of the
$\tau$ would mimic a dilepton signal but with two missed
neutrinos washing out the expected $e (2j)$ peak.   

It has to be borne in mind that the excess seen in the $(2e)(2j)$       
mode is not matched in the $(2\mu)(2j)$ data. This would have to be     
interpreted as an indication that the right-handed neutrino associated  
with the muon, $N_\mu$, is significantly heavier than $N_e$ and so      
its production in $W_R$-decay suffers a large kinematic suppression.    
Further, the coupling of $N_e$ to $\mu$ has to be small, i.e.,
$|V_{N_e\mu}| \ll 1$.

Interpretation of the excess in terms of $N_e$, a Majorana
neutrino, would lead to the expectation of roughly an equal
number of like-sign and unlike-sign dilepton events.  There are
fourteen events in the excess region in the CMS data in the
$(2e)(2j)$ channel, of which only one has charged leptons of the
same sign. In a similar analysis by the ATLAS collaboration no
like-sign events are found \cite{atlike}. One way around this is to
assume that two degenerate right-handed neutrinos together form a
pseudo-Dirac state in which case the like-sign events are
suppressed \cite{other2}.

\begin{figure}[tbh] 
\begin{center} 
    \includegraphics[width=.47\textwidth]{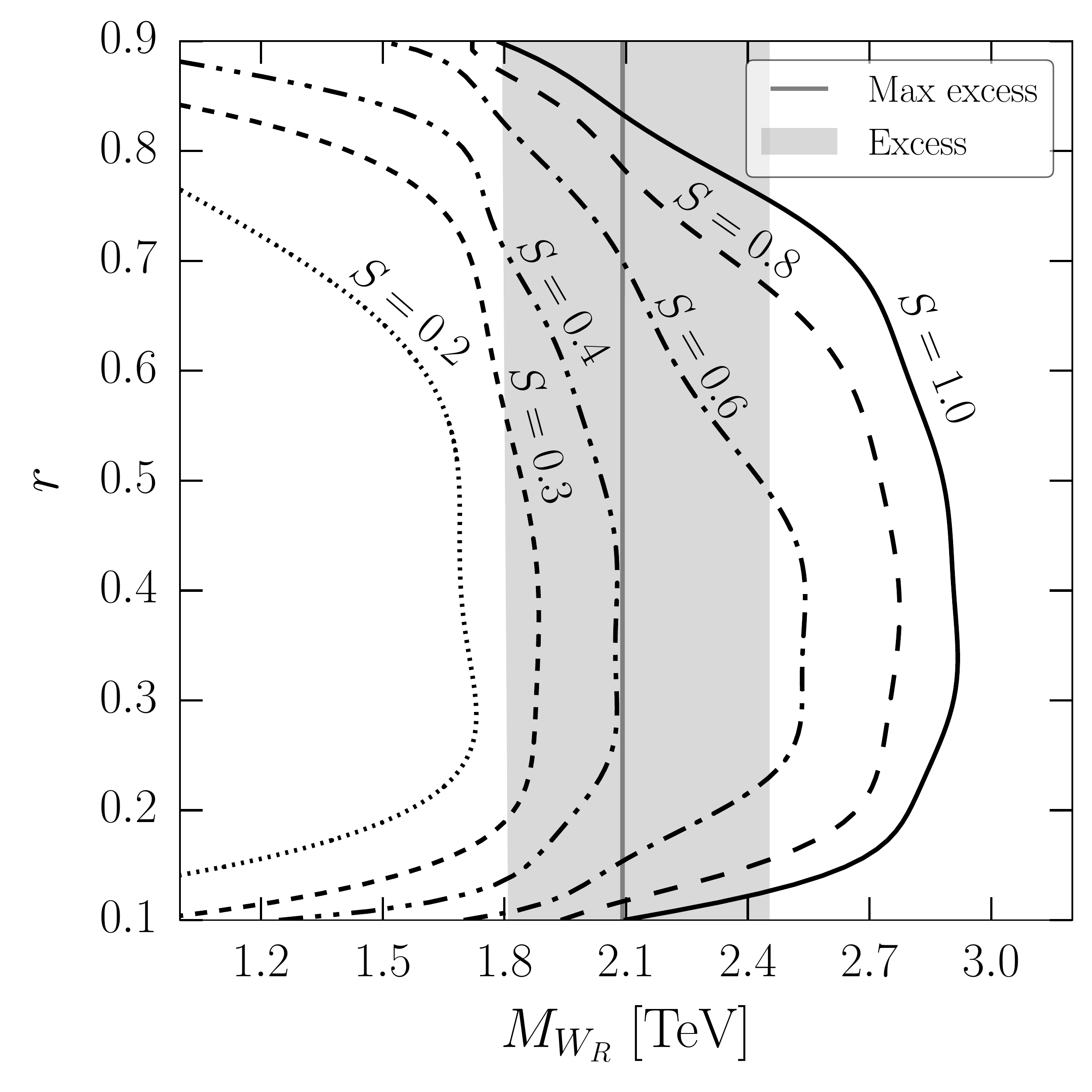}
\caption{\em The shaded region demarcates the range of $M_{W_R}$
for which the CMS data exceed twice the SM expectation. The
maximum excess is on the vertical straight line. The curves
parametrised by $S \equiv \eta |V_{N_ee}|^2$ denote the $(r,
M_{W_R})$ contours for which the prediction of the LRS model is
compatible with the observation.} 
\label{obs1} 
\end{center} 
\end{figure} 

In Fig. \ref{obs1} we place the excess
observed by CMS in this channel -- the shaded region in the
$M_{W_R} - r$ plane -- in comparison with the LRS model predictions.  This
excess is maximum along the vertical line.
The expectations from the Left-Right
Symmetric model ($\sigma BR_T$) depend on $S^2 = \eta^2
|V_{Ne}|^4$ and $r = M_{N_e}/M_{W_R}$. The  dashed curves in the
figure, identified by the values of $S$, trace the points in the
$(r - M_{W_R})$ plane for which the LRS expectations equal
$\sigma BR_O$. To put the plot in context note that CMS has
stressed \cite{cms} that with $\eta = 1$ and $V_{Ne} = 1$ --
i.e., $S = 1$ -- the LRS model signal for $r = 0.5$ is
inconsistent with the excess. This is borne out from Fig.
\ref{obs1} which indicates that for the $S=1$ contour, the
$M_{W_R}$ corresponding to $r=0.5$ lies outside the excess
region. Consistency of the excess in the data with the LRS model
can be accomplished in three ways. Firstly, if $r =
M_{N_e}/M_{W_R}$ is larger than 0.5 the LRS model signal will be
reduced. Indeed, with $r >$ 0.75 the LRS model is consistent with
the excess even with $S = 1$. Alternatively, if $\eta$ or
$V_{Ne}$ is less than unity, then too the signal will be less,
the suppression being determined by $S^2$. In Fig. \ref{obs1} it
can be seen that for $r$ = 0.5 the excess is consistent with the
model for $0.3\lesssim S\lesssim 0.6$.  The upper limit has been
pointed out in \cite{utpal} and \cite{othergut}. What we essentially
find is that there are large sets of values of $r$, $\eta$ and 
$V_{Ne}$ for which the LRS expectation is consistent with the excess.

\begin{figure}[tb] 
\begin{center} 
    \includegraphics[width=.40\textwidth]{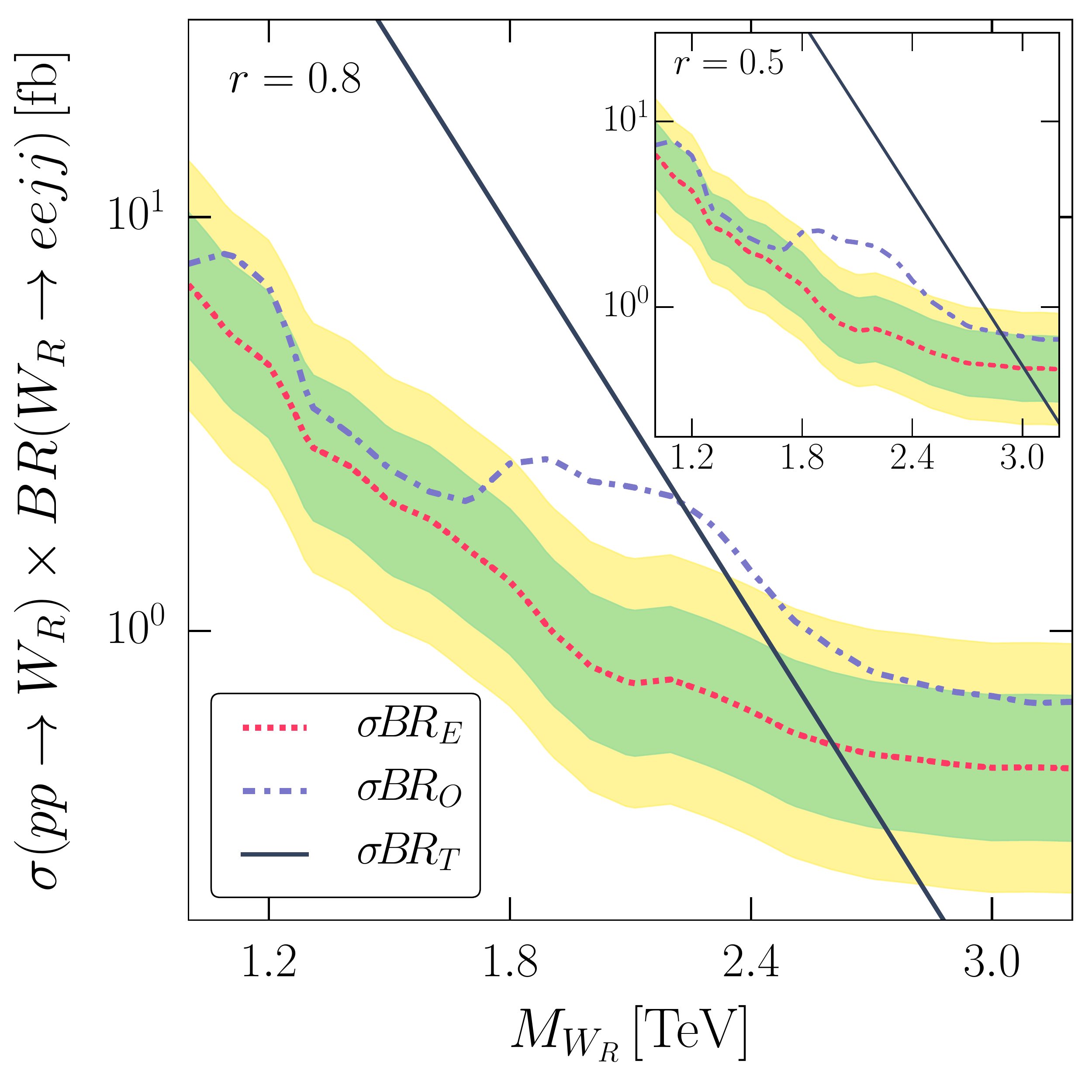}
\hskip 20pt
    \includegraphics[width=.40\textwidth]{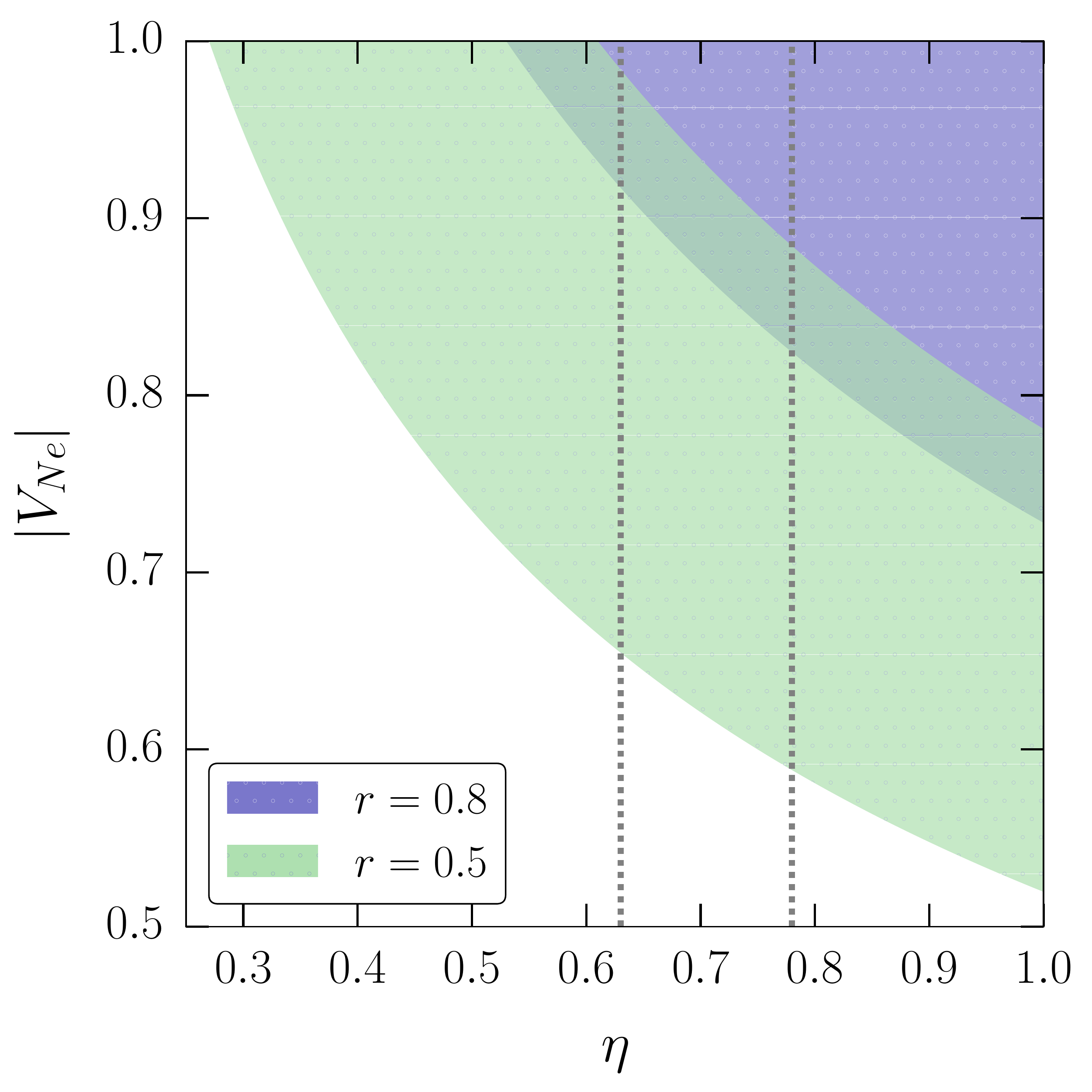}
\caption {\em Left: The CMS data compared with the LRS model
predictions for $r$ = 0.8 keeping $g_R/g_L$ =1 and $V_{Ne}$ = 1.
Inset: $r$ = 0.5. Right: $\eta$ and $V_{Ne}$ that fits the excess in the
CMS data for $r$ = 0.5 and $r$=0.8. Only the region between the
two vertical lines is permitted in $SO(10)$ GUTs.}
\label{obs1a} 
\end{center} 
\end{figure}

Fig. \ref{obs1} contains information in a somewhat condensed
form. In the spirit of the path chosen by the CMS collaboration,
we use the exclusion data and plot in the left panel of Fig.
\ref{obs1a} $\sigma BR_E$ (red dotted curve) and $\sigma BR_O$
(blue dashed curve) as functions of $M_{W_R}$ for the fixed value
of $r$ = 0.8.  The prediction of the LRS model with $\eta =
g_R/g_L = 1$ and $V_{N_ee} = 1$ is the black solid straight line.
Also shown are the $\pm$50\% (green, dark) and  $\pm$100\%
(yellow, light) bands of the expected cross section. In the inset
the same results are presented but for $r$ = 0.5.  Notice that
for $r = 0.8$ the LRS model expectation passes right through the
maximum of the excess while for $r = 0.5$ it entirely misses the
excess region.

The right panel of Fig. \ref{obs1a} utilises a complementary way
of displaying the region in the LRS model parameter space
consistent with the result.  Here the area in the $\eta -
V_{N_ee}$ plane that fits the CMS excess region is shown shaded
for two values of $r$ = 0.8  (violet, dark) and 0.5 (green,
light).  Note that there is an overlap region.
It is worth stressing that, as in the left panel, for
$r$ = 0.8 the model is consistent with the data even for $\eta
=1$ and $V_{N_ee}$ = 1.  For $r = 0.5$ a suppression through the
factor $S = \eta |V_{Ne}|^2$ is required to bring the model in
harmony with the data. If the $W_R$ with a mass ${\cal O}$(TeV)
arises from the $SO(10)$ GUT model we discuss below then $\eta$
must lie within the two vertical lines.

\section{$SO(10)$ Grand Unification}

$SO(10)$ is an attractive candidate for a unified theory
\cite{SO10a, SO10b}    
as it is the simplest Lie group which includes all the SM fermions      
and a right-handed neutrino of one generation in a single irreducible   
representation. We do not include any exotic fermions in the model 
and deal with three generations.

There are a vast number of models characterised by different
intermediate symmetries which have $SO(10)$ as the unifying
group.  In that respect $SO(10)$ is more of an umbrella term,
incorporating these different models with alternate
symmetry-breaking routes, scalar structures, and physics
consequences.  What is important for this work is that $SO(10)$
has the Pati-Salam symmetry ($\mathcal{G}_{PS}$) as a
subgroup\footnote{$SO(10)$ can break into ${\mathcal G}_{SM}$ through two
distinct routes, one through an intermediate $SU(5)$ with no
left-right symmetry and another through the PS stage.  A longer
proton decay lifetime $\tau_p$ than predicted by minimal $SU(5)$
and the ease of incorporation of seesaw neutrino masses give the
second option a slight preference.} and includes the discrete  D-parity
\cite{Dparity} which enforces left-right parity, $g_L = g_R$. The
Left-Right Symmetric group is embedded in $\mathcal{G}_{PS}$.
Thus, having reviewed the CMS result in terms of the LRS model,
both with and without left-right parity, the obvious next step is
to look at it through the lenses of the Pati-Salam partial
unified and $SO(10)$ grand unified theories.

In this section we summarize the features of $SO(10)$ GUTs which are    
relevant for our subsequent discussions. We consider the        
non-supersymmetric version of this theory.                              

\subsection{Symmetry breaking}

\begin{table}[h!]
\begin{center}
\begin{tabular}{|c|c|c|c|c|c|c|}\hline
Symmetry & $SO(10)$ & D-Parity & $SU(4)_C$ & $SU(2)_R$ &
$U(1)_R \times U(1)_{B-L}$ & $SU(2)_L \times U(1)_Y$ \\ \hline
Breaking Scale & $M_U$ & $M_D$ & $M_C$ & $M_R$ & $M_0$ & $M_{Z}$ \\ \hline     
\end{tabular}
\end{center}
\caption{\em The different scales at which subgroups of $SO(10)$ get broken.}
\label{t:scales}
\end{table}

\begin{figure}[bth]
    \centering
    \includegraphics[width=0.6\textwidth]{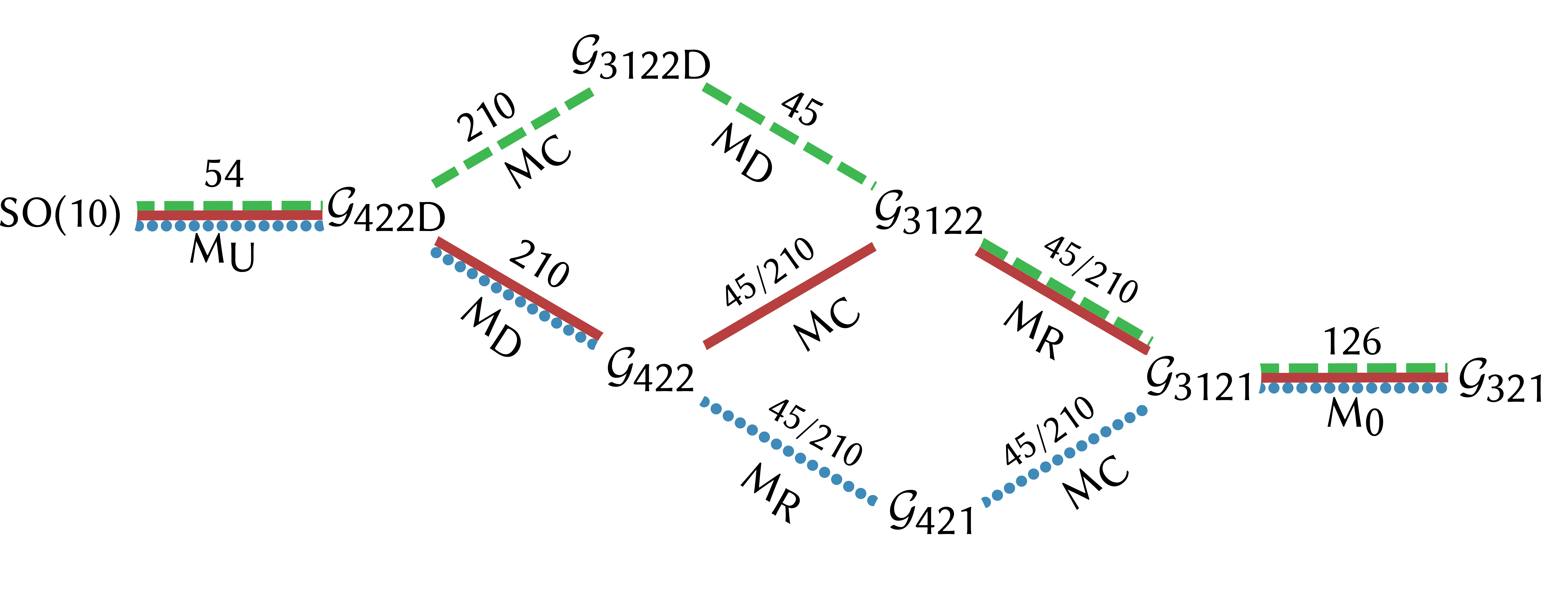}
    \caption{\em Symmetry breaking routes of $SO(10)$ distinguished
by the order of breaking of $SU(2)_R$, $SU(4)_C$, and D-parity.
The  scalar multiplets responsible for symmetry breaking
at every stage have been indicated. Only the DCR (red solid)
route can accommodate the light $W_R$ scanario.}
    \label{fig:routes}
\end{figure}

The different ways in which $SO(10)$ GUT can step-wise break to
the SM are graphically represented in Fig. \ref{fig:routes}.  The
intermediate energy scales of various stages of symmetry breaking
will be denoted according to Table \ref{t:scales}. Among these,
$M_D \geq M_R \geq M_0 > M_Z$ always. In order to systematically
study the different ways in which $SO(10)$ can descend to the SM,
we first classify them into \emph{routes} based on the order of
symmetry breaking. We will call the route with $M_C\geq M_D \geq
M_R$, CDR (Green, Dashed), the one with $M_D \geq M_C \geq M_R$, DCR
(Red, Solid), and  another,  DRC (Blue, Dotted), with $M_D\geq
M_R \geq M_C$. Thus there are three alternate routes with a
maximum number of four intermediate stages. Among the
intermediate stages the first and the last, namely, $SU(4)_C
\times (SU(2)_L\times SU(2)_R)_D$ ($\equiv {\mathcal G}_{422D}$)
and $SU(3)_C \times U(1)_{B-L} \times SU(2)_L\times U(1)_R $
($\equiv {\mathcal G}_{3121}$), are common to all routes.  The
other possible intermediate symmetries, in this notation, are
${\mathcal G}_{422}$, ${\mathcal G}_{421}$, ${\mathcal
G}_{3122D}$, and ${\mathcal G}_{3122}$ (see Fig.
\ref{fig:routes}).  All models of $SO(10)$ symmetry breaking
(symmetry-breaking \emph{chains}) are thus defined by the route
it belongs to and the Higgs multiplets that it includes.  Figure
\ref{fig:routes}  shows the {\em maximum-step chains}
(chains with maximum number of intermediate symmetries) of each
route. Other chains are essentially subcases of these with
multiple symmetries breaking at the same scale. This can be
achieved if multiple Higgs sub-multiplets gain vacuum expectation
value ({\em vev}) at the same scale or if a single sub-multiplet
breaks more than one symmetry.

\subsection{Scalar structure and the Extended Survival Hypothesis       
(ESH)}                                                                  

The gauge bosons in the model and their masses are determined by the    
symmetry group and its sequential breaking to the SM. The fermions come 
in three generations in each of which there are the SM quarks and    
leptons and a right-handed neutrino. Thus it is only the scalar sector  
which retains a degree of flexibility.

The generation of quark and lepton masses requires a
\underline{10} of $SO(10)$ while the see-saw mechanism for
neutrino masses relies on a \underline{126}. Their decompositions
under the PS group are\footnote{We will use the
notation $[\phi_4,\phi_L,\phi_R]$ to specify the behaviour of $SO(10)$
submultiplets under the Pati-Salam  symmetry.}:
\begin{equation}
10 = [1,2,2] + [6,1,1]   \;\;,
\label{h10}
\end{equation}
and
\begin{equation}
126 = [6,1,1] + [15,2,2] + [10,3,1] +  [\overline{10},1,3]  \;\;.
\label{h126}
\end{equation}
These scalars also have important roles in gauge symmetry
breakings. The {\em vev} of the \underline{10}, which is ${\cal
O}$($M_Z$), breaks the standard model $SU(2)_L \times U(1)_Y$
symmetry while the \underline{126} is responsible for the
breaking of $U(1)_{B-L} \times U(1)_R$ at the scale $M_0$.

In a grand unified theory masses of fermions in the same multiplet
are related. In particular, the \underline{10} of $SO(10)$ implies
$M_d = M_l^\dagger$, where $M_d$ is the mass matrix of $d$-type quarks
and $M_l$ that of the charged leptons. Though these relations are
valid only at the scale of unification and at lower energies
corrections have to be included, even then they are not in
consistency with the measured masses. One way to address this
issue is to invoke  a [15,2,2] submultiplet which is present in
the \underline{126} and the \underline{120} of $SO(10)$ 
to bring the masses closer to their actual values
\cite{fmass1, fmass2}. 

Two other $SO(10)$ representations which turn out to be useful
for symmetry breaking and whose submultiplet structure will be
important are the \underline{45} and \underline{210}. Under the
PS group they consist of:
\begin{equation}
45 = [15,1,1] + [6,2,2] + [1,3,1] +  [1,1,3]  \;\;, 
\label{h45}
\end{equation}

\begin{equation}
210 = [1,1,1] + [15,1,1] + [6,2,2]  + [15,3,1] +  [15,1,3] +
[10,2,2]  + [\overline{10},2,2] \;\;.
\label{h210}
\end{equation}

Even using these limited $SO(10)$ multiplets\footnote{A
$\underline{54}$ is used for the first step of GUT symmetry
breaking. It does not affect the RG running of the couplings.},
there remains a variety of options for the scalar submultiplets
that can be used for the different stages of symmetry breaking
indicated in Fig. \ref{fig:routes}. They affect the unification
and intermediate energy scales through their role in the
evolution of gauge couplings. We make two restrictions:
(a) Only renormalisable terms will be kept in the
$SO(10)$-symmetric lagrangian\footnote{This excludes, for
example, using scalar \underline{16}-plets to mimic the $SO(10)$
\underline{126}  for neutrino mass through effective dimension-5
terms in the Lagrangian.}, and (b) The Extended Survival
Hypothesis, which is a consequence of minimal fine-tuning, is
taken to be valid.

According to ESH \cite{ESH1, ESH2}, at any intermediate
energy scale only those scalar submultiplets (under the unbroken
symmetry at that stage)  which are required to spontaneously
break a symmetry at that  or any lower energy  remain massless.
All other submultiplets become massive.  Because the normal
expectation of scalar masses is to be at the highest energy scale
the extended survival hypothesis posits the minimal number of
fine-tunings in the scalar sector.

With these guiding principles we now turn to the scalar
multiplets that are employed for the descent of $SO(10)$ to the
SM. The first (at $M_U$) and last (at $M_0$) stages of the
symmetry breaking in Fig. \ref{fig:routes}, which are common to
all alternate channels, utilise a \underline{54}-plet and a
\underline{126}-plet of scalar fields, respectively. D-parity is
broken through the {\em vev} of D-odd scalars. There is a D-odd
PS singlet in the \underline{210} of $SO(10)$ which can be
utilised for the DCR or DRC symmetry-breaking route. For the CDR
route a D-odd LRS model singlet in the [15,1,1] of \underline{45} is
useful.

\subsection{Renormalisation Group Equations}

The one-loop RG evolution for the coupling $\alpha^g(\mu)$
corresponding to a gauge symmetry $g$ can be
written as:
\begin{equation}
    \frac{1}{\alpha^{g}(\mu_i)} = \frac{1}{\alpha^{g}(\mu_j)} +
  \frac{b^{g}_{ji}}{2\pi}\ln\left ( \frac{\mu_j}{\mu_i}\right ).
    \label{eq:rg1}
\end{equation}
$b^g_{ji}$ is the coefficient of the $\beta$-function
between the scales
$\mu_i$ and $\mu_j$:
\begin{equation}
    b^g = -\frac{11}{3}N_g + \frac{2}{3}\sum_{F}T(F_g) d(F_{g'}) n_G +
    \frac{1}{6}\sum_{S} \delta_S
    T(S_g) d(S_{g'})  ,
\end{equation}
where the three terms are contributions from gauge bosons, chiral
fermions, and scalars respectively. $N_g$ is the quadratic
Casimir corresponding to the particular symmetry group $g$, $N_g$
is 0 for $U(1)$ and $N$ for $SU(N)$. $T(F_g)$ and $d(F_g)$ are
the index and the dimension of the representation of the chiral
fermion multiplet $F$ under the group $g$ and the sum is over all
fermion multiplets of one generation. $n_G$ is the number of
fermion generations, 3 in our case. Similarly $T(S_g)$ and
$d(S_g)$ are the index and the dimension of the representation
$S_g$ of the scalar $S$ under $g$. $\delta_S$ takes the value 1
or 2 depending on whether the scalar representation is real or
complex.

It is worth noting that $b^g$ is positive for $U(1)$ subgroups
and negative\footnote{Contributions from large scalar multiplets
can make the beta-function positive. This does happen for
$SU(2)_R$ in the example we discuss later.} for $SU(n)$.
Therefore, $U(1)$ couplings grow with increasing energy while
$SU(n)$ couplings decrease.

For ease of use, we will rewrite eq. (\ref{eq:rg1}) as:
\begin{equation}
    w^g_i = w^g_j + \frac{1}{2\pi}b^g_{ji}\Delta_{ji} \;\;, 
\label{eq:rg_redef}
\end{equation}
where $w^g_i\equiv\frac{1}{\alpha^{g}(\mu_i)}$ and $\Delta_{ji}\equiv
\ln\left ( \frac{\mu_j}{\mu_i}\right )$.

If the symmetry $g$ is broken to $g'$ at the scale $\mu_i$ then
the coupling constant matching condition is simply $w^g_i =
w^{g'}_i$ unless two groups combine to yield a residual symmetry.
As an example of the latter, for $U(1)_Y$ of the standard model,
resulting from a linear combination of $U(1)_R$ and $U(1)_{B-L}$
at the scale $M_0$, one has:
\begin{equation}
    w^Y_0=\frac{3}{5}w^R_0 + \frac{2}{5}w^{B-L}_0 \;\;.
\label{u11}
\end{equation}

Matching all the couplings at the boundaries and imposing the
unification condition one arrives at three equations:
\begin{eqnarray}
    w^3_{Z}    &=&  w_U + \frac{1}{2\pi}\sum_i b^C_{i,i-1}\Delta_{i,i-1} \;\;,
\nonumber\\ 
    w^{2L}_{Z} &=&  w_U + \frac{1}{2\pi}\sum_i b^{2L}_{i,i-1}\Delta_{i,i-1} \;\;,
\nonumber \\ 
    w^{Y}_{Z}  &=&  w_U + \frac{3}{5}\frac{1}{2\pi}\sum_i b^{1R}_{i,i-1}\Delta_{i,i-1} +
    \frac{2}{5}\frac{1}{2\pi}\sum_i b^{B-L}_{i,i-1}\Delta_{i,i-1} \;\;,
\label{eq:system}
\end{eqnarray}
where $w_U$ is the reciprocal of the coupling strength at unification. $i$
runs from the unification scale to $M_0$.
$C$ stands for $SU(3)_C$ or $SU(4)_C$ depending on the energy
scale $\mu$. Similarly,  $1R$ ($B-L$) in the last equation represents
$U(1)_R$ or $SU(2)_R$ ($U(1)_{B-L}$ or $SU(4)_C$). 

The left-hand-sides  of the three equations in (\ref{eq:system})
are the inputs fixed by experiments. The equations  are linear in
$w_U$ and $\ln(\mu_i)$ -- the logarithms of the mass-scales.
There are 2+$m$ variables: $m$, the number of scales intermediate
to $M_U$ and $M_{Z}$, $w_U$, the magnitude of the coupling  at
unification, and the GUT scale $M_U$ itself. Thus, an $SO(10)$
chain with one intermediate scale ($m = 1$) is a determined
system while those with more steps are underdetermined.


\section{Low energy expectations from unification}

In the LRS model the energy scales of symmetry breaking can be
freely chosen to be consistent with the low energy data. Once
embedded in GUTs one must also verify that such choices of
intermediate scales are consistent with perturbative unification
of the couplings at sub-Planck energies and check their
implications for other symmetry-breaking scales. In this section
we look at the restrictions imposed by coupling unification
on $\eta$ and the other left-right symmetric model parameters. 

$SO(10)$ can descend to the SM through a maximum of four
intermediate stages (Fig. \ref{fig:routes}). Such four-step
symmetry breakings are underdetermined.  Accordingly, one is
permitted to choose the scale $M_R$ in the TeV range, as required
by the CMS data, and to check the consistency of the equations.
$M_0$ is always below $M_R$ and thus keeping the latter at a few
TeV  sets the former to an even lower value.

\subsection{Pati-Salam partial unification}

The PS symmetry with D-parity, ${\mathcal G}_{422D}$, is  a
common intermediate stage for all the $SO(10)$ symmetry-breaking options.
When D-parity is intact, this model has two-independent
couplings, namely, $g_{4C}$ and $g_{2L} = g_{2R} = g_2$, which
achieve equality at the grand unification scale $M_U$.  In the
DCR  route the Pati-Salam ${\mathcal G}_{422}$ survives at the
next step but D-parity no longer holds.  In contrast, for the CDR
chain the PS symmetry is broken before D-parity. Needless to say,
so long as D-parity remains unbroken $\eta = 1$.

In  Pati-Salam partial unification one has a set of three
equations similar to eq. (\ref{eq:system}) sans the constraint of
grand unification.   In place of an inverse GUT coupling $w_U$
one gets two separate couplings -- $w^4_C$ $(= w_C^{B-L} =
w_C^3)$ at $M_C$ and $w^2_D$ ($ = w^{2R}_D = w^{2L}_D)$ at $M_D$.
Thus, the two variables -- the GUT coupling and the GUT scale --
are replaced by the $SU(4)_C$ unification coupling and a D-parity
symmetric $SU(2)$  coupling.  In the following sections we will
look at the results that arise from RG evolution for both PS
partial unification and $SO(10)$ grand unification.

\subsection{Left-right symmetry and unification}

The scalar field contributions to gauge coupling evolution play a
significant role in achieving coupling unification while keeping
a low $M_R$. This has led to a plethora of models where scalar
fields have been incorporated in the theory solely for this
purpose.  This is not the path that we choose.  Indeed, the
scalar fields which we {\em do} include become indispensible in
some cases. For example, a subcase which one might imagine from
Fig. \ref{fig:routes} will have $M_R = M_0$.  The one-step
symmetry breaking of ${\mathcal G}_{3122} \rightarrow {\mathcal
G}_{321}$ can be realized through the  {\em vev} of just a
[$\overline{10}$, 1, 3] $\subset$ \underline{126}, dispensing off
the submultiplet which breaks $SU(2)_R \rightarrow U(1)_R$.
However, without this latter contribution the coupling constants
no longer unify. So, $M_R = M_0 \sim {\cal O}$(TeV) cannot be
accommodated without at least the scalar  multiplets that we keep.

As mentioned earlier, the three key ingredients in interpreting
the CMS result are the ratio between the left- and right-handed
gauge couplings, $\eta$, the Majorana mass of the right-handed
electron neutrino, $M_{N_e}$, and the right-handed leptonic
mixing $V_{N_ll}$.  The Majorana mass of the right-handed
neutrino of the $l$-th flavour, in the TeV range, is obtained
through the Yukawa coupling $Y_{126}^l$. The mass is proportional
to the $\Delta L = 2$ {\em vev}, $v_{126}$, of the $(1, -2, 1,1) \subset [\overline{10}, 1, 3] \subset$
\underline{126}. The latter also breaks the ${\mathcal G}_{3121}$
symmetry. Hence, one has $M_{N_l} \sim (Y_{126}^l/g_{B-L}) M_0$.
The Yukawa coupling, $Y_{126}^l$, can be chosen to obtain a
desired value of $M_{N_l}$ without affecting other physics. Thus
the choice of $r = M_N/M_{W_R}$ is decoupled from the analysis of
coupling unification.

The mixing in the right-handed lepton sector -- $V_{N_ll}$ --  
is the second relevant quantity in this analysis. It is determined by   
the generation structure of the Yukawa matrix. Since this does not   
affect the evolution of couplings, which is the focus, our analysis     
does not impose any restriction on the choice of this mixing.

The relative strength of the right-handed coupling {\em vis-\`{a}-vis}             
the left-handed one at the $SU(2)_R$-breaking scale --               
$\eta=\frac{g_R}{g_L}$ -- is, however, intimately related to the RG       
running of the gauge couplings.       
\begin{eqnarray}
    w^{2R}_R &=& \frac{1}{\eta^2} w^{2L}_R \;\;. 
\end{eqnarray}
The magnitude of $\eta$ will vary for symmetry-breaking chains
depending on the scalar content of the theory and the energy
scales at which different symmetries break.  Nonetheless, the
minimum value that can be attained by $\eta$ is almost
independent of the way in which $SO(10)$ or $\mathcal{G}_{PS}$
descends to the standard model, as we now discuss.
Firstly the requirement that $M_R$ is $\mathcal O$(TeV) and $M_0$
even lower, keeps them close to each other and the two are not too
far from $M_Z$ either. The other feature, noted earlier, is that $U(1)$
couplings increase as the energy scale $\mu$ increases while
$SU(n)$ couplings do the opposite.

One starts from eq. (\ref{u11}) which
relates the $U(1)$ couplings when the symmetry breaking
${\mathcal G}_{3121} \rightarrow {\mathcal G}_{321}$ occurs at
$M_0$. Obviously,
\begin{equation}
w^{1R}_0 > w^{1R}_R = w^{2R}_R = \left(\frac{1}{\eta^2}\right)w^{2L}_R   
\;\;,
\label{u12}
\end{equation}
and from $w^{B-L}_C = w^{3C}_C$ 
\begin{equation}
w^{B-L}_0 > w^{B-L}_R > w^{3C}_R \;\;.   
\label{u13}
\end{equation}
From eq. (\ref{u11}) together with eqs.  (\ref{u12}) and (\ref{u13}) one has
\begin{equation}
\eta^2 > \frac{3 w^{2L}_R}{5 w^Y_0 - 2  w^{3C}_R} \simeq 
\frac{3 w^{2L}_0}{5 w^Y_0 - 2  w^{3C}_0} \;\;.   
\label{etalim1}
\end{equation}

The inequality in the first step in eq. (\ref{u12}) is due to the
evolution of $w^{1R}$ from $M_0$ to $M_R$. Since these two energy
scales are both in the TeV range this effect is not large.  A
similar reasoning is also valid for the first inequality in eq.
(\ref{u13}) but the second could be much more  substantial. Using
the current values of the low energy couplings\footnote{We use
$\alpha_3$ = 0.1185(6), $\sin^2\theta_W$ =  0.23126(5), and
$\alpha =$ 1/127.916 at $\mu = M_Z$ \cite{PDG}.} and
extrapolating them to $\mu = M_0$ one gets 
\begin{equation}
\eta_{min}\sim 0.59 \;\;.
\label{e:etamin}
\end{equation}
We stress that eq. (\ref{e:etamin}) is an artefact of the LRS
model so long as there is a merging of the $U(1)_{B-L}$ with
$SU(3)_C$, and so  is valid for both PS (partial) and $SO(10)$
(grand) unification. However, this is a limit in principle,
accomplishing it will depend on the details of symmetry breaking
and the scalar content of the theory. We have previously seen
 that if $r = 0.5$ the CMS result is compatible with the LRS
model for $S$ as low as $\sim 0.25$. From the preceding
discussion we see that $S$ lower than $\sim 0.59$ cannot be
attained by $\eta$ alone.

\section{The three routes of $SO(10)$ symmetry breaking}

In this section we consider one by one the three routes depicted
in Fig. \ref{fig:routes} by which $SO(10)$ can descend to the SM.
We focus on the scalar fields that are required and the
intermediate energy scales involved.  We use one-loop
renormalisation group equations here but have checked that 
two-loop effects -- on which we comment later on --  do not
change the results drastically. Since the equations are usually
underdetermined, motivated by the CMS data, we will keep 4 TeV
$\leq M_R \leq $ 10 TeV and 1 TeV $\leq M_0 \leq $ 4 TeV for the
chains of descent.


\subsection{The DRC route}

Restricting $M_R$ to the TeV range automatically eliminates the
DRC route  (Blue dotted in Fig. \ref{fig:routes}) -- $SU(4)_C$
breaking {\em after} $M_R$ -- because then the leptoquark gauge
bosons of $SU(4)_C$ achieve a mass of the TeV order. Light
leptoquarks below 10$^6$ GeV are forbidden from rare decays of
strange mesons, such as $K_L \rightarrow \mu e$
\cite{PS, Kmue1, Kmue2}. 
The DRC route of symmetry breaking is thus not compatible with the CMS result.


\subsection{The CDR  route}
\label{secCDR}

\begin{table}[hbt]
\begin{center}
\begin{tabular}{|c|c|c|c|c|c|c|}
\hline
$SO(10)$&Symmetry&\multicolumn{5}{|c|}{Scalars contributing to
RG}\\ \cline{3-7} 
repn. & breaking & $M_Z \leftrightarrow M_0$ & $M_0 \leftrightarrow M_R$ &
 $M_R \leftrightarrow M_D$
& $M_D \leftrightarrow M_C$ & $M_C \leftrightarrow M_U$\\
 & &   ${\mathcal G}_{321}$ &  ${\mathcal G}_{3121}$ &
 ${\mathcal G}_{3122}$ & ${\mathcal G}_{3122D}$ 
& ${\mathcal G}_{422D}$   \\ \hline
{\bf 10} & ${\mathcal G}_{321} \rightarrow EM$ & (1,2,$\pm1$) & 
(1,0,2,$\pm\frac{1}{2}$) &
(1,0,2,2) & (1,0,2,2)$_+$ & [1,2,2]$_+$\\  
& & & & & &\\
{\bf 126} & ${\mathcal G}_{3121} \rightarrow {\mathcal G}_{321}$ &
-  & (1,-2,1,1) & (1,-2,1,3) & (1,-2,1,3)$_+$ & 
[$\overline{10}$,1,3]$_+$\\  
& & -  & - & - & (1,2,3,1)$_+$ & [10,3,1]$_+$\\  
 & & & & & & \\
{\bf 210} & ${\mathcal G}_{3122} \rightarrow {\mathcal G}_{3121}$ &
-  & - & (1,0,1,3) & (1,0,1,3)$_+$ & [15,1,3]$_+$\\  
& & -  & - & - & (1,0,3,1)$_+$ & [15,3,1]$_+$\\  
 & & & & & & \\
{\bf 210} & ${\mathcal G}_{3122D} \rightarrow {\mathcal G}_{3122}$ &
-  & - & - & (1,0,1,1)$_-$ & [1,1,1]$_-$\\  
 & & & & & & \\
{\bf 210} & ${\mathcal G}_{422D} \rightarrow {\mathcal G}_{3122D}$ &
-  & - & - & - & [15,1,1]$_+$\\  
\hline
\end {tabular}\\
\caption{\em Scalar fields considered when the ordering of
symmetry-breaking scales is $M_C \geq M_D \geq M_R$.  The
submultiplets contributing to the RG evolution at different
stages  according to the ESH are shown. D-parity ($\pm$) is
indicated as a subscript.  }
\label{t:eshCDR}
\end {center}
\end{table}

With all intermediate stages distinct, for this route (Green
dashed in Fig. \ref{fig:routes}) one has:
\begin{equation}
    SO(10)\xrightarrow[54]{M_U} {\mathcal G}_{422D}
\xrightarrow[210]{M_C} {\mathcal G}_{3122D} 
    \xrightarrow[210]{M_D} {\mathcal G}_{3122} \xrightarrow[210]{M_R}
    {\mathcal G}_{3121} \xrightarrow[126]{M_0} {\mathcal G}_{321} \;\;.
\end{equation}
The scalar submultiplets responsible for the symmetry breaking
are shown in Table \ref{t:eshCDR}.  An alternative to the above
would be  to break ${\mathcal G}_{3122} \rightarrow {\mathcal
G}_{3121}$ using a [1,1,3] $\subset$ \underline{45} in place of
the $[15,1,3] \subset$ \underline{210}. We also comment about
this option.

In order to proceed with an elaboration of the consequences
associated with this route it is helpful to list the one-loop
beta-function coefficients for the stages $M_R \leftrightarrow
M_D$ and $M_D \leftrightarrow M_C$.  Including the contributions
from the scalars in Table \ref{t:eshCDR}, fermions, and gauge
bosons one finds from eq. (\ref{eq:rg1})
\begin{eqnarray}
b^3_{DR} &=& -7 \;,\;b^{B-L}_{DR} = \frac{11}{2} \;,\;b^{2L}_{DR} = -3 \;,\;
b^{2R}_{DR} = -2 \;,\;  \nonumber \\
b^{3}_{CD} &=& -7 \;,\;b^{B-L}_{CD} = 7 \;,\;b^{2L}_{CD} = -2
\;,\;b^{2R}_{CD} = -2 \;.\;
\label{betaCDR}
\end{eqnarray}
The $SU(2)_L$ and $SU(2)_R$ couplings evolve from $M_R$ to become
equal at $M_D$. This requires ($\Delta_{AB} = \ln \frac{M_A}{M_B}$):
\begin{equation}
w^{2L}_R - w^{2R}_R = \frac{1}{2\pi} \left\{(b^{2L}_{DR} -
b^{2R}_{DR}) \Delta_{DR} \right\} \;\;.
\label{CDR2L2R}  
\end{equation}
Similarly the $SU(3)_C$ and $U(1)_{B-L}$ couplings become equal at
$M_C$, i.e., 
\begin{equation}
w^{3}_R - w^{B-L}_R = \frac{1}{2\pi} \left\{(b^{3}_{DR} -
b^{B-L}_{DR}) \Delta_{DR}  + 
(b^{3}_{CD} - b^{B-L}_{CD}) \Delta_{CD} \right\} \;\;.
\label{CDR3C1BL}  
\end{equation}
The left-hand-sides of eqs. (\ref{CDR2L2R}) and (\ref{CDR3C1BL})
are given in terms of the various couplings at $M_R$. Since $M_R
\sim {\cal O}$(TeV) and the RG evolution is logarithmic in energy
it is not a bad approximation to assume that they do not
change significantly from $M_Z$ to $M_R$, i.e., $w^i_R \simeq
w^i_O \simeq w^i_Z$. Then recalling eq. (\ref{u11}) which relates
$w^Y_0$ with $w^R_0$ and $w^{B-L}_0$ one can obtain:
\begin{equation}
\small{
3 w^{2L}_Z + 2 w^3_Z - 5 w^Y_Z \simeq \frac{1}{2\pi} \left\{
\left[3(b^{2L}_{DR} - b^{2R}_{DR}) + 2 (b^{3}_{DR} -
b^{B-L}_{DR})\right] \Delta_{DR}  + 2 (b^{3}_{CD} - b^{B-L}_{CD})
\Delta_{CD} \right\} .
\label{CDRreln}
}
\end{equation}

Using the beta-function coefficients from  eq. (\ref{betaCDR}), one can
reexpress eq. (\ref{CDRreln}) as:
\begin{equation}
3 w^{2L}_Z + 2 w^3_Z - 5 w^Y_Z \simeq \frac{1}{2\pi} \left\{
28 ~\Delta_{CR}  \right\} \;\;.
\label{CDRreln2}
\end{equation} 
Notice that $M_D$ has dropped out. Further, the
low energy values of $\alpha, \alpha_s$ and $\sin^2\theta_W$
\cite{PDG} then imply $M_C \sim 10^{18} M_R$, i.e., way beyond
the Planck scale. The low energy SM paramters are now quite
well-measured and offer no escape route from this impasse.
Two-loop contributions also do not change the situation
drastically. We have checked that if one breaks ${\mathcal
G}_{3122} \rightarrow {\mathcal G}_{3121}$ through a [1,1,3]
$\subset$ \underline{45} rather than the $[15,1,3] \subset$
\underline{210} (see Table \ref{t:eshCDR}), the change is in the
evolution of the couplings in the $M_C \leftrightarrow M_U$
sector which does not affect this conclusion.

The above analysis does not resort to the constraint of grand
unification at all. The results hold for PS partial unification
as well. So, the CDR route of descent also has to be abandoned
for $M_R \sim {\cal O}$(TeV).

\subsection{The DCR route}
\label{secDCR}

\begin{table}[hbt]
\begin{center}
\begin{tabular}{|c|c|c|c|c|c|c|}
\hline
$SO(10)$&Symmetry&\multicolumn{5}{|c|}{Scalars contributing to
RG}\\ \cline{3-7} 
repn. & breaking & $M_Z \leftrightarrow M_0$ & $M_0 \leftrightarrow M_R$ &
 $M_R \leftrightarrow M_C$
& $M_C \leftrightarrow M_D$ & $M_D \leftrightarrow M_U$\\
 & &   ${\mathcal G}_{321}$ &  ${\mathcal G}_{3121}$ &
${\mathcal G}_{3122}$ &  ${\mathcal G}_{422}$ 
&  ${\mathcal G}_{422D}$   \\ \hline
{\bf 10} & ${\mathcal G}_{321} \rightarrow EM$ & (1,2,$\pm1$) & 
(1,0,2,$\pm\frac{1}{2}$) &
(1,0,2,2) & [1,2,2] & [1,2,2]$_+$\\  
& & & & & &\\
{\bf 126} & ${\mathcal G}_{3121} \rightarrow {\mathcal G}_{321}$ &
-  & (1,-2,1,1) & (1,-2,1,3) & [$\overline{10}$,1,3] & 
[$\overline{10}$,1,3]$_+$\\  
& & -  & - & - & - & [10,3,1]$_+$\\  
 & & & & & & \\
{\bf 210} & ${\mathcal G}_{3122} \rightarrow {\mathcal G}_{3121}$ &
-  & - & (1,0,1,3) & [15,1,3] &  [15,1,3]$_+$\\  
& & -  & - & - & - & [15,3,1]$_+$\\  
 & & & & & & \\
{\bf 210} & ${\mathcal G}_{422} \rightarrow {\mathcal G}_{3122}$ &
-  & - & - & [15,1,1] & [15,1,1]$_+$\\  
 & & & & & & \\
{\bf 210} & ${\mathcal G}_{422D} \rightarrow {\mathcal G}_{422}$ &
-  & - & - & - & [1,1,1]$_-$\\  
\hline
\end {tabular}\\
\caption{\em Scalar fields considered when the ordering of
symmetry-breaking scales is $M_D \geq M_C \geq M_R$.  The
submultiplets contributing to the RG evolution at different
stages  according to the ESH are shown. D-parity ($\pm$) is
indicated as a subscript. }
\label{t:eshDCR}
\end {center}
\end{table}

After having eliminated the other alternatives, the only
remaining route of descent has the mass ordering $M_D \geq M_C \geq
M_R$ (Red solid in Fig. \ref{fig:routes}).  Keeping all possible
intermediate stages separate from each other this corresponds to:

\begin{equation}
    SO(10)\xrightarrow[54]{M_U} {\mathcal G}_{422D} 
    \xrightarrow[210]{M_D} {\mathcal G}_{422}
    \xrightarrow[210]{M_C} {\mathcal G}_{3122}
    \xrightarrow[210]{M_R} {\mathcal
     G}_{3121}\xrightarrow[126]{M_0} {\mathcal G}_{321}  \;\;.
\label{DCRchain}
\end{equation}

In the above we have indicated the $SO(10)$ multiplets which
contribute to symmetry breaking at every stage.   The scalar
submultiplets which contribute to the RG equations as dictated by
ESH are shown in Table \ref{t:eshDCR}.         
There is, however, an alternative which relies on a \underline{45}      
of $SO(10)$ whose contents under the Pati-Salam group are given in      
eq. (\ref{h45}). $SU(2)_R$ can be broken by the (1,0,1,3) $\subset       
[1,1,3] \subset$ \underline{45} replacing the [15,1,3] $\subset$        
\underline{210}. In fact, the $SU(4)_C$ breaking [15,1,1] is also       
present in the \underline{45}. However, one cannot entirely dispense    
with the \underline{210} because the [1,1,1]$_-$ in it has no analog in 
the \underline{45}.                                                    

Denoting by $h_D, h_C, h_R$ the $SO(10)$ scalar multiplets
responsible for the breaking of D-Parity, $SU(4)_C$, and
$SU(2)_R$ respectively, we therefore have the following
alternatives:  \{$h_D, h_C, h_R$\} can be
\{210,45,45\}, \{210,45,210\}, \{210,210,45\} and \{210,210,210\}.      
Of these, the first employs the lowest dimensional scalar multiplets    
required to break symmetries at each scale while the last one uses the  
least number of $SO(10)$ scalar multiplets. Using \underline{45} or     
\underline{210} for $h_C$ makes no difference in the physics since in   
both cases a [15,1,1] Pati-Salam submultiplet is used. The distinction  
is relevant only in the choice of $h_R$.                                

The one-loop beta-function coefficients for the couplings in the
$M_R \leftrightarrow M_C$ and $M_C \leftrightarrow M_D$ energy
ranges obtained using eq. (\ref{eq:rg1}) and the scalars in Table
\ref{t:eshDCR} are:
\begin{eqnarray}
b^3_{CR} &=& -7 \;,\;b^{B-L}_{CR} = \frac{11}{2} \;,\;b^{2L}_{CR} = -3 \;,\;
b^{2R}_{CR} = -2 \;,\;  \nonumber \\
b^{4}_{DC} &=& -5 \;,\;b^{2L}_{DC} = -3
\;,\;b^{2R}_{DC} = \frac{26}{3} \;.\;
\label{betaDCR}
\end{eqnarray}
The $SU(3)_C$ and $U(1)_{B-L}$ couplings evolve to become equal
at $M_C$. Thus
\begin{equation}
w^{3}_R - w^{B-L}_R = \frac{1}{2\pi} \left\{(b^{3}_{CR} -
b^{B-L}_{CR}) \Delta_{CR}   \right\} \;\;.
\label{DCR3C1BL}  
\end{equation}
Matching of the $SU(2)_L$ and $SU(2)_R$ couplings at $M_D$ implies: 
\begin{equation}
w^{2L}_R - w^{2R}_R = \frac{1}{2\pi} \left\{(b^{2L}_{CR} -
b^{2R}_{CR}) \Delta_{CR} 
+ (b^{2L}_{DC} - b^{2R}_{DC}) \Delta_{DC} \right\} \;\;.
\label{DCR2L2R}  
\end{equation}
As before, we use the 
approximation  $w^i_R \simeq w^i_O \simeq w^i_Z$ and combine eqs.
(\ref{DCR3C1BL}) and (\ref{DCR2L2R}) to get: 
\begin{equation}
\small{
3 w^{2L}_Z + 2 w^3_Z - 5 w^Y_Z \simeq \frac{1}{2\pi} \left\{ \left[3(b^{2L}_{CR} -
b^{2R}_{CR}) + 2 (b^{3}_{CR} - b^{B-L}_{CR})\right] \Delta_{CR} + 
3 (b^{2L}_{DC} - b^{2R}_{DC}) \Delta_{DC}  \right\} .
\label{DCRreln}
}
\end{equation}
A special limit of the DCR route is when $M_D = M_C$, i.e.,
$\Delta_{DC} = \ln \frac{M_D}{M_C} = 0$. In this
limiting case there is no distinction between this route and the
CDR one. Indeed, setting $M_D = M_C$ in eq. (\ref{DCRreln}) and
substituting  the beta-function coefficients from
eq. (\ref{betaDCR}) one exactly
reproduces (\ref{CDRreln2}) which places the solution in an
unacceptable energy regime.

That one should nonetheless expect acceptable solutions can be
surmised from the fact that eq. (\ref{DCRreln}) implies
\begin{eqnarray}
\frac{d \ln M_C}{d \ln M_D} &=&  \frac{3 (b^{2L}_{DC} - b^{2R}_{DC})}
{[3(b^{2L}_{DC} - b^{2R}_{DC} - b^{2L}_{CR} + b^{2R}_{CR}) 
- 2 (b^{3}_{CR} - b^{B-L}_{CR})] } \nonumber \\
&=& 5  \;\;,
\end{eqnarray}
where in the last step we have used eq. (\ref{betaDCR}). This
indicates that $M_C$ changes faster than $M_D$ and so starting
from the $M_D = M_C$ limit solutions in the DCR route with the
symmetry breaking scales below $M_{Planck}$
are feasible. Replacing the \underline{210} in $h_R$ by a
\underline{45} reduces $b^{2R}_{DC}$ so much that unification of
couplings is no longer possible. In the
next section we present the allowed soultions in detail.



\section{$SO(10)$ unification with $M_R\sim\mathcal{O}$(TeV)}

In the previous section we have seen that of the three routes of
symmetry breaking accessible to $SO(10)$, DRC is trivially
eliminated when the twin requirements $M_R \sim {\cal O}$(TeV)
and  $M_R > M_C$ are imposed. We also indicated that  for
the CDR route with the minimal scalar content  and following the
extended survival hypothesis the requirement $M_R \sim
{\cal O}$(TeV) implies $M_C > M_{Planck}$.  The only route that can
accommodate $M_R \sim {\cal O}$(TeV) is DCR.

To simplify the discussion, in eq.  (\ref{DCRreln}) we have
ignored the running of the couplings between $M_Z$ and $M_R$. In
obtaining the results presented in this section we have not used
such an approximation.  We have, however, not included the
effect of mixing of $U(1)_R \times U(1)_{B-L}$ under RG
evolution from $M_0$ to $M_R$. As these two scales are close to
each other, both in the few TeV range, the impact of the mixing
will not be large.

\subsection{Pati-Salam partial unification for the maximum-step case}

The maximum-step symmetry-breaking DCR route has been
given in eq. (\ref{DCRchain}). Before turning to $SO(10)$ we
briefly remark about Pati-Salam partial unification within this
route.  Because there are four steps of
symmetry breaking this is an underdetermined system. For this
work, $M_R$ is restricted to be in the $\mathcal O$(TeV) range.
The scale $M_C$ is taken as the other input in the analysis. At
the one-loop level the results can be analytically calculated
using the beta-function coefficients in eq. (\ref{betaDCR}). The
steps can be identified from eqs. (\ref{DCR2L2R}) and
(\ref{DCRreln}). The latter determines $M_D$ once $M_C$ is
chosen. $\eta$ is then fixed using eq. (\ref{DCR2L2R}). 

For example, for $M_C = 10^6$ GeV one gets $\eta = 0.63$ when
$M_R =$ 5 TeV. Within the Pati-Salam model the upper limit of
$M_D$ is set by  $M_{Planck}$. We find that in
such a limit one has $M_C = 10^{17.6}$ GeV and $\eta =$ 0.87 for
$M_R $ = 5 TeV.

\subsection{Coupling unification for the maximum-step case}

For $SO(10)$ grand unification one must find the energy scale at
which the common $SU(2)_{L,R}$ coupling beyond $M_D$ equals the
$SU(4)_C$ coupling, i.e.,   $g_2 = g_{4C}$. This limits the upper
 bound of $M_C$ compared to the Pati-Salam partial unification. 

\begin{figure}[thb]
    \centering
\includegraphics[width=0.45\textwidth]{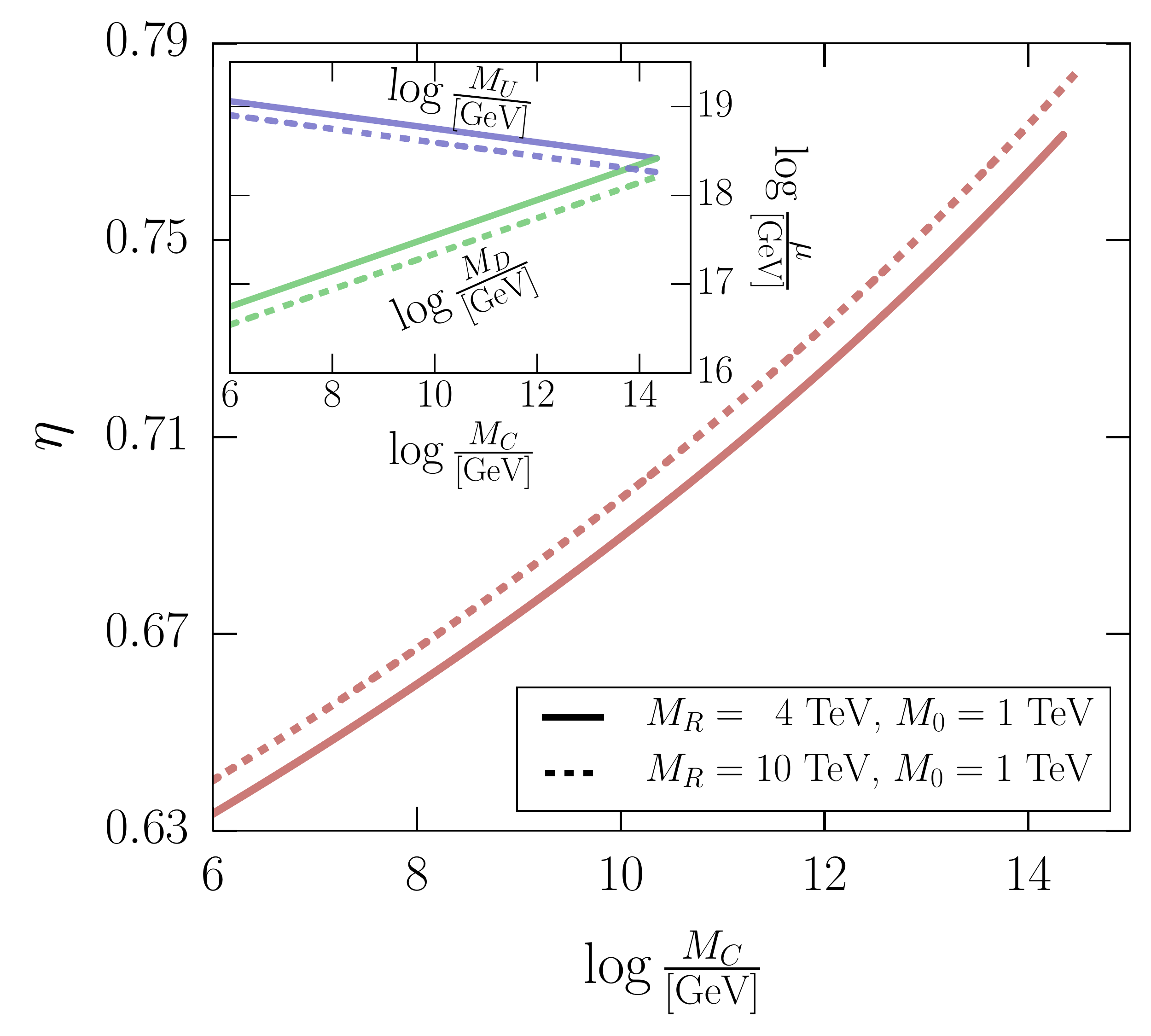}
        \includegraphics[width=0.45\textwidth]{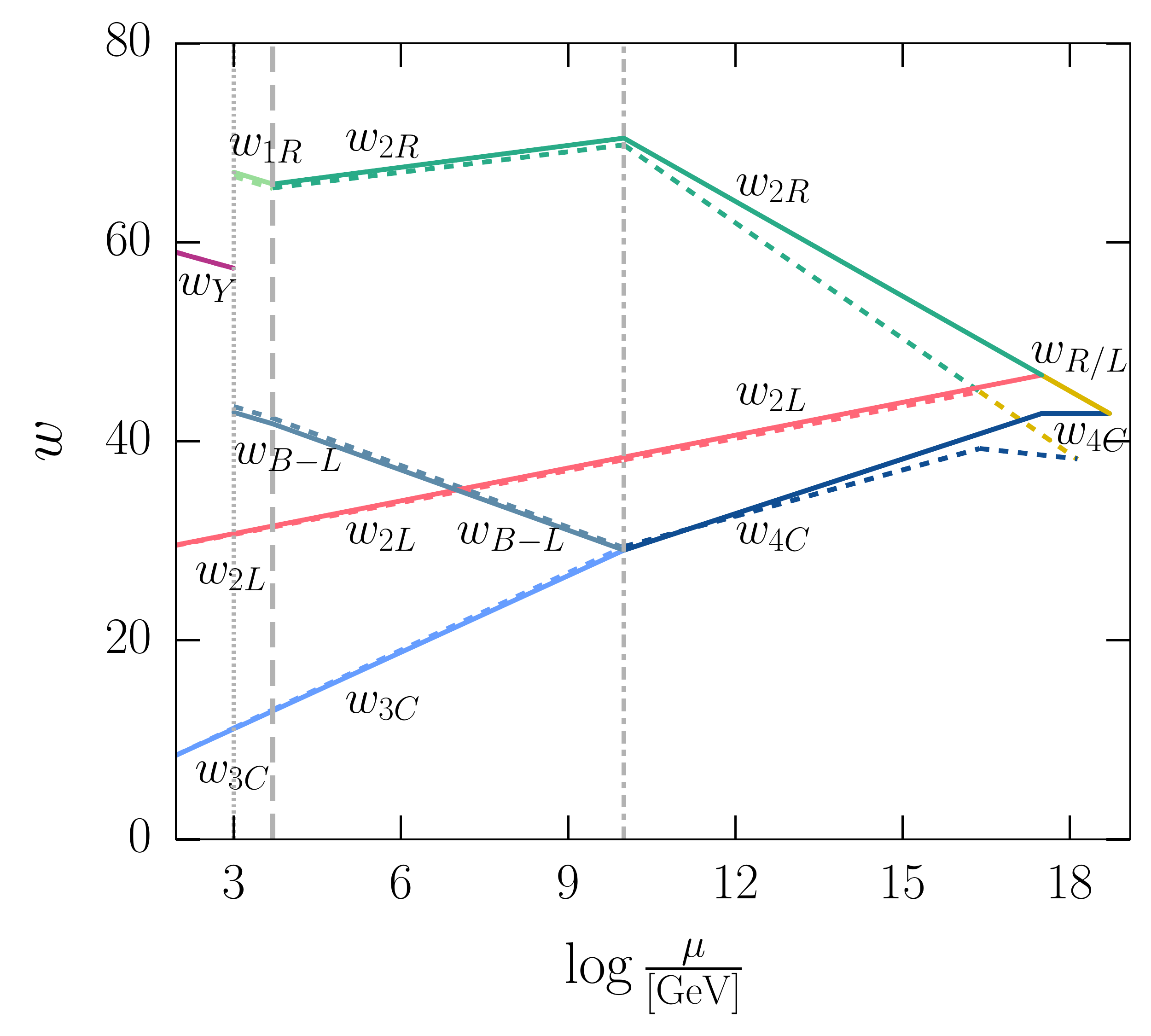}
\caption{\em   Left: $\eta$ is plotted as a function of $M_C$ for
the DCR chain.  In the inset the behaviour of $M_D$ and $M_U$ are
shown.  The two curves in each case correspond to $M_R$ = 4
(solid) and 10 TeV (dashed). In both cases $M_0$ = 1 TeV is
taken. Right: Behaviour of the gauge couplings for the DCR chain
of $SO(10)$ GUT with $M_0$ = 1 TeV, $M_R$ = 5 TeV and $M_C$ =
$10^{10}$ GeV. The solid (dashed) lines correspond to one-loop
(two-loop) evolution of couplings.
The scalar fields are as in Table \ref{t:eshDCR}.}
\label{fig:DCR_mass}
\end{figure}

In the left panel of Fig. \ref{fig:DCR_mass} we plot $\eta$ as a
function of $M_C$. In the inset is shown the behaviour of $M_U$
and $M_D$ as functions of $M_C$. Due to the unification
constraint, the upper limits of $M_C$, $M_D$ and $\eta$ all
decrease from the respective values which were obtained in the PS
case.  The lowest value of $\eta$ turns out to be $\sim$ 0.63.
Notice that a lower value of $M_C$ is associated with a higher
$M_U$, which must not exceed $M_{Planck}$.   $M_C$ is also
bounded from below by the experimental limits on flavour-changing
transitions such as $K_L \rightarrow \mu e$.  It is this that
determines the lowest admissible $M_C$, in general. From the
inset it is seen that although $M_U$ increases as $M_C$
decreases, it remains below $M_{Planck}$ so long as $M_C > 10^6$
GeV. As $M_C$ increases $M_D$ increases as well and the point
where it meets the decreasing $M_U$ determines the upper limit of
$M_C$. For every plot the ranges consistent with 4 TeV $\leq M_R
\leq$ 10 TeV are between the two curves, the solid one indicating
the $M_R$ = 4 TeV end.  The results are almost insensitive to
the choice of $M_0$ between 1 TeV and $M_R$. Note that
irrespective of the scale of $SU(4)_C$ breaking, $M_D$ always
remains above $10^{16}$ GeV. The unification coupling constant,
$w_U$, varies between 38.4 and 47.6 and thus perturbativity
remains valid throughout.


The behaviour of the coupling constants as a function of the
energy scale for a typical case of $M_0$ = 1 TeV, $M_R$ = 5 TeV
and $M_C = 10^{10}$ GeV are shown (solid lines) in the right
panel of Fig. \ref{fig:DCR_mass}.  Note that due to the
contributions of large scalar multiplets to the $\beta$-functions
the coupling $g_{2R}$ grows beyond $M_C$.  Although this chain is
suited to our needs, the unification scale is close to the Planck
scale for $M_R\sim\mathcal{O}$(TeV). Thus, if $M_{W_R} \sim {\cal
O}$(TeV) then it is unlikely that ongoing proton decay
experiments \cite{SK} will observe a signal.  This is a
consequence of our adhering to the principle of minimality of
Higgs scalars.  One can lower $M_U$ by including scalars
redundant to symmetry breaking. 

We have set the lower limit of $M_C$ at $10^6$ GeV from the
limits on rare meson decays such as $K_L \rightarrow \mu e$ or
$B_{d,s}\rightarrow \mu e$. The current limit on the branching
ratio for the former process is $Br(K_L \rightarrow \mu^\pm
e^\mp) < 4.7 \times 10^{-12}$ at 90\% CL \cite{PDG} which
translates to $M_C \gtrsim 10^6$ GeV. LHCb has set the tightest
bounds on the latter processes.  They find (again at 90\% CL)
\cite{LHCBBmue} $Br(B_d^0 \rightarrow \mu^\pm e^\mp) < 2.8 \times
10^{-9}$ and $Br(B_s^0 \rightarrow \mu^\pm e^\mp) < 1.1 \times
10^{-8}$ which yield a weaker limit on $M_C$. It can be expected
that these bounds will be strengthened when the results from the
newer runs of LHC appear.  In addition,  $n - \bar{n}$
oscillations can be mediated through coloured scalars belonging
to the $[\overline{10}, 1, 3] \subset$ \underline{126} which also
acquire mass at the scale of $M_C$.  The current experimental
limit, $\tau_{n-\bar{n}} \geq 2.7 \times 10^{8}$ s \cite{nnSK} at
90\% CL, also translates to $M_C \gtrsim 10^6$ GeV.  Therefore,
improvements in the measurement of the above-noted rare meson
decays and $n - \bar{n}$ oscillations open the possibility of
probing, at least in part, the GUT options that can accommodate a
TeV-scale $W_R$.

\subsection{The $M_D = M_U$ case}

There are a number of daughter chains of the DCR route with two
symmetries breaking at the same scale. Of these,  the choice
$M_C = M_R$, resulting in a common point of the DCR and DRC
routes, violates the lower bound on $M_C$ from flavour changing
processes since $M_R \sim {\cal O}$(TeV). As noted in the
previous section, another alternative, namely, $M_D = M_C$,
which is a point shared by the DCR and CDR routes, occurs at an
energy beyond the Planck scale. The only remaining possibility is
$M_D = M_U$.

The upper limit on $M_C$ is set by the requirement $M_D=M_U$.
This  happens when D-parity is broken at the GUT scale by a
[1,1,1]$_- \subset$ \underline{210}. We thus have
\begin{eqnarray}
    SO(10)\xrightarrow[210]{M_U=M_D} {\mathcal G}_{422}
    \xrightarrow[210]{M_C} {\mathcal G}_{3122}
    \xrightarrow[210]{M_R}
    {\mathcal G}_{3121} \xrightarrow[126]{M_0} {\mathcal G}_{321}\;\;.
\label{DCR_1d}
\end{eqnarray}

From the inset in the left panel of Fig. \ref{fig:DCR_mass} it is
seen that for $M_C\sim 2\times 10^{14}$ GeV one has $M_D=M_U$. As
this chain has three intermediate steps, there are no free
parameters after setting $M_0$ and $M_R$. The coupling at
unification, $w_U$, comes to be
around 47.6, and $\eta$, as can be seen from the left panel of Fig.
\ref{fig:DCR_mass}, is near 0.78.  An interesting aspect of
this chain is that it is minimal in the number of scalar
multiplets used.

\subsection{Two-loop comparison}


\begin{figure}[h!]
    \centering
    \includegraphics[width=0.45\textwidth]{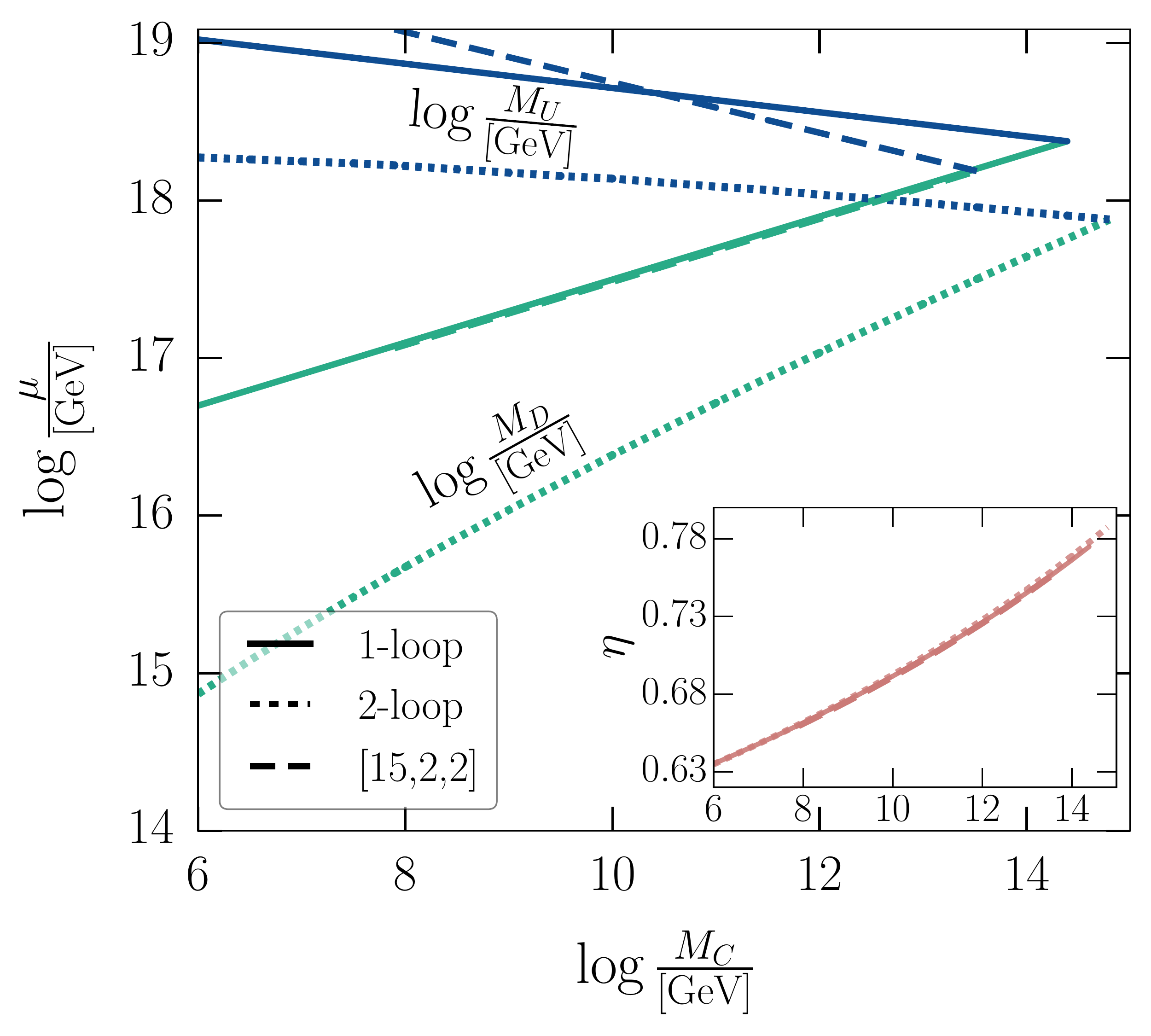}
    \caption{\em A comparison of the results for one-loop (solid
lines) and two-loop (dotted line) running.  The evolution of
$M_U$  and $M_D$  with $M_C$ is
displayed. Also shown is the case of one-loop running when a
[15,2,2] scalar multiplet is added (dashed lines). In the inset
the variation of $\eta$ is presented.}
\label{fig:2loop}
\end{figure}

The discussion till now, based on RG evolution using
one-loop $\beta$-functions, was amenable to an analytical
examination.  Our aim was to reason our way through
different $SO(10)$ symmetry-breaking options in search of chains
which can accommodate a TeV range $M_R$.  Now, after finding a
specific pattern for which $M_R \sim \mathcal{O}($TeV) is
tenable, we indicate the size of two-loop effects for this
chain. In the right panel of Fig. \ref{fig:DCR_mass} the
evolution of the couplings for a typical choice of $M_C =
10^{10}$ GeV, $M_R = 10$ TeV and $M_0 = 1$ TeV are indicated by
the dashed lines.  It is seen that the essential physics is
largely unaltered though there is some change in the various
energy scales.

The chain in (\ref{DCRchain}) contains many scalar fields until
$SU(4)_C$ breaking, contributing heavily to the two-loop
$\beta$-coefficients \cite{2loop1, 2loop2}. These are responsible
for some departures from the one-loop results. We have presented
the one-loop results for the DCR route keeping $M_C$ as an input
parameter. In Fig. \ref{fig:2loop} we compare the one-loop (solid
lines) and two-loop (dotted lines) results for $M_U$ and $M_D$ as
a function of $M_C$.   The deviation for both scales increases
with decreasing $M_C$. This is intuitive because with lower $M_C$
$SU(4)_C$ remains a good symmetry for a larger energy range over
which the two-loop contributions are effective. In the inset a
similar comparison is made for $\eta$.   It is noteworthy  that
$\eta$ remains largely unaffected.

\subsection{Additional scalars: Two examples}

If fermion masses are generated through scalars which belong to
the minimal set required for symmetry breaking then mass
relationships do not reflect observed values.  One needs to add
at least one extra scalar multiplet -- which is light, couples to
fermions, and develops a {\em vev} at the electroweak scale -- to
get realistic mass ratios.  In $SO(10)$, fermions reside in the
\underline{16} representation and scalars transforming only as
\underline{120}, \underline{126}, and \underline{10} can have
 Yukawa couplings, since:
\begin{equation}
    16\times 16 = 10 + 120 + 126  \;\;. \label{eq:bidoublet}
\end{equation}
In  $SO(10)$ GUTs improved fermion mass relations can be obtained
\cite{fmass2} using the PS submultiplet [15,2,2] $\subset$
\underline{126} in addition to the [1,2,2] $\subset$
\underline{10}.  The natural scale for the extra scalar
submultiplet would have been at the GUT scale and extra
fine-tuning is necessary to keep it at the electroweak scale.

We have examined the behaviour of gauge coupling evolutions for the DCR
case including the additional [15,2,2] submultiplet to check if
the TeV range $M_R$ still remains viable.  In Fig.
\ref{fig:2loop}  the variation of $M_D$ and $M_U$ with $M_C$ when
an extra [15,2,2] is included are shown (dashed lines). The
effect on $\eta$ (shown in the inset) is negligible. The
important change is that the permitted lowest $M_C$ is more
restricted as $M_U$ tends rapidly towards $M_{Planck}$. 

It is clear that the scale $M_D$ is governed  by the difference
in the $\beta$-coefficients of $SU(2)_R$ and $SU(2)_L$. 
Submultiplets such as [15,2,2], which contribute symmetrically to the
$\beta$-coefficients of the left- and right-handed $SU(2)$
groups, will not affect the difference  and so do
not change the D-parity breaking scale.  For the reasoning 
$\eta$ is not affected as well.

Another alternative we examined pertains to the impact of
additional scalars necessary to produce two right-handed
neutrinos nearly degenerate in mass but with opposite
CP-properties. As noted in Sec. \ref{sec:WR} this can explain the
lack of like-sign dilepton events in the CMS data
\cite{other2}. A right-handed neutrino mass matrix of the appropriate
nature can be generated \cite{dibosonTH1} by a scalar multiplet
$\chi \equiv$ (1,2,1) under the $SU(2)_L\times SU(2)_R\times
U(1)_{B-L}$ symmetry. In an $SO(10)$ GUT $\chi$ is a member of
the \underline{16} representation and must develop a {\em vev}
at the scale $M_0$. We have examined the impact of adding a
\underline{16}-plet on $\eta = g_R/g_L$, the symmetry breaking
scales, and the coupling at the unification point. We find no
serious effect on any of these, the allowed range of $M_C$ is
about one order of magnitude larger, and for any $M_C$, the
unification scale, $M_U$, and the D-parity-breaking scale, $M_D$,
are both somewhat lowered. $\eta$ is essentially unaffected.

\section{Summary and Conclusions}

The observation by the CMS collaboration of a 2.8$\sigma$ excess
in the $(2e)(2j)$ channel around 2.1 TeV can be interpreted as a
preliminary indication of the production of a right-handed gauge
boson $W_R$. Within the left-right symmetric model the excess
identifies specific values of $\eta = g_R/g_L$, $r =
M_N/M_{W_R}$, and $V_{N_ee}$. We stress that even with $g_R = g_L$
and $V_{N_ee}$ = 1 the data can be accommodated by an appropriate
choice of $r$.

We explore what the CMS result implies if the left-right
symmetric model is embedded in an $SO(10)$ GUT. $\eta \neq 1$ is
a consequence of the breaking of left-right D-parity. We find
that a $W_R$ in the few TeV range very tightly restricts the
possible routes of descent of the GUT to the standard model. The
only sequence of symmetry breaking which is permitted is $M_D >
M_C > M_R > M_0$ with a D-parity breaking scale $\geq 10^{16}$ GeV. All
other orderings of  symmetry breaking  are excluded. 
Breaking of left-right discrete parity at such a high scale
pushes $g_L$ and $g_R$ apart and one finds $0.64 \leq \eta \leq
0.78$.  The unification scale, $M_U$,  has to be as high as $\sim
10^{18}$ GeV so that it is very unlikely that proton decay will
be seen in the ongoing experiments. The $SU(4)_C$-breaking scale,
$M_C$, can be as low as $10^6$ GeV, which may be probed by rare
decays such as $K_L \rightarrow \mu e$ and $B_{d,s} \rightarrow
\mu e$ or $n - \bar{n}$ oscillations. In Table [\ref{tab:res}]
we summarise the essence of the allowed GUT solutions. We have
assumed that no extra scalar multiplets are included beyond those
needed for symmetry breaking and invoked the Extended Survival
Hypothesis to identify scalar submultiplet masses.

\begin{table}[h!]
\begin{center}
    \small
    \begin{tabular}[]{ccccccc}
        \hline
        Sr&Intermediate Symmetries&\multicolumn{3}{c}{Mass
    Scales $\left(\log \frac{\mu}{\rm [GeV]} \right)$}&$w_U$&$\eta$\\
        No.&  & $M_U$ & $M_D$ & $M_C$ & &  \\
        \hline
        & & & & & &  \\
        1&${\mathcal G}_{422D}\rightarrow
        {\mathcal G}_{422}\rightarrow
        {\mathcal G}_{3122}\rightarrow {\mathcal G}_{3121}$
        &  19.02 - 18.38 &  16.70 - 18.38
        & 6.00 - 14.39 &  38.41 - 47.63  & 0.64 - 0.78\\
        & & & & & &  \\
        2& 2-loop &  18.27 - 17.88 &  14.87 - 17.88
        & 6.00 - 14.80 &  29.54 - 46.66  & 0.64 - 0.79\\
        & & & & & &  \\
        3& Added [15,2,2] scalar & $M_{Planck}$ - 18.19 &  17.06 - 18.19
        & 7.90 - 13.52 &  18.59 - 37.52  & 0.66 - 0.76\\
        & & & & & &  \\
        \hline
    \end{tabular}
    \caption{\em The $SO(10)$ symmetry-breaking chains consistent with $M_R$
= 5 TeV and $M_0$ = 1 TeV. The intermediate symmetries and the associated
mass-scales are shown.}
    \label{tab:res}
\end{center}
\end{table}
The ATLAS collaboration has  indicated \cite{diboson}  an
enhancement around 2.1 TeV in the di-boson -- $ZZ$ and $WZ$ --
channels in their 8 TeV data. Our interpretation of the excess in
the $(ee)(jj)$ channel can be extended to
include the branching ratio of $W_R$ to di-boson states
\cite{rizzo, keung2}. One can arrange to accommodate both these findings if
$r \simeq 1$ and $V_{N_ee} < 1$. Such a combined analysis is beyond the
scope of this work and will be reported elsewhere.   Results
along somewhat similar directions can be found in the literature
\cite{utpal3}.  It has  also been shown that interpretations of
the di-boson observations are possible if the LRS model is
embellished with the addition of some extra fermionic states
\cite{dibosonTH1, dibosonTH2}.

{\bf Acknowledgements:} TB acknowledges a Junior Research
Fellowship from UGC, India.  AR is partially funded by  the
Department of Science and Technology Grant No. SR/S2/JCB-14/2009.

\end{document}